\documentclass{article}

\usepackage{graphicx}

\usepackage{enumitem}
\usepackage[margin=.75in]{geometry}

\newcommand{\FIG}{Fig.~}
\newcommand{\FIGS}{Figs.~}
\newcommand{\SEC}{Sect.~}
\newcommand{\SECS}{Sects.~}

\title{Global and local synchrony of coupled neurons in  small-world networks}
\author{Naoki Masuda$^1$ and Kazuyuki Aihara$^{2,3}$\\
$^1$ Faculty of Engineering, Yokohama National University, Yokohama, Japan\\
$^2$ Graduate School of Frontier Sciences, The University of Tokyo, Tokyo, Japan\\
$^3$ ERATO, Japan Science and Technology Agency, Kawaguchi, Japan}
\date{}

\begin{document}
\maketitle

\begin{abstract}
Synchronous firing of neurons is thought to play important functional roles such as feature binding and switching of cognitive states. Although synchronization has mainly been investigated using model neurons with simple connection topology so far, real neural networks have more complex structures. Here we examine behavior of pulse-coupled leaky integrate-and-fire neurons with various network structures. We first show that the dispersion of the number of connections for neurons influences dynamical behavior even if other major topological statistics are kept fixed. The rewiring probability parameter representing the randomness of networks bridges two spatially opposite frameworks: precise local synchrony and rough global synchrony. Finally, cooperation of the global connections and the local clustering property, which is prominent in small-world networks, reinforces synchrony of distant neuronal groups receiving coherent inputs.
\end{abstract}

\section{Introduction}\label{intro}

Synchronization of neurons is thought to play crucial roles in
information processing in the brain.  For example, the neurons sharing
a common preferred stimulus fire synchronously when stimulated
\cite{Gray}.  Switching between synchrony and asynchrony also occurs
in response to modality changes during tasks \cite{Steinmetz}.
Synchronously firing neurons may form a dynamical cell assembly to
bind features such as visual objects \cite[p. 145, 150]{Domany},
\cite{Fujii,Konig,Vondermalsburg86,Wang95_IEEE}.  Synchronous cell
assemblies are robust against noise and can provide, by using the
temporal dimension, a solution to the feature binding problem that
involves combinatorial explosion \cite[p. 142, 145]{Domany}.  In
accordance, the mechanisms of synchronization and also those of
desynchronization and clustered states have been explored using model
neural networks with the complete couplings \cite{Vanvreeswijk96},
random sparse couplings \cite{Golomb00,Wang96}, and local couplings
\cite{Konig,Terman,Wang95_IEEE}.

However, the network topology considered in these works is too simple
to give satisfactory answers to many basic questions.
For example, all-to-all and random networks
blur the information stored in local modules or individual neurons
\cite{Terman}.  These architectures cannot explain the emergence of
local clusters owing to relatively dense local couplings
\cite{Sompolinsky91} while dynamic clusters formed by stimulus
application or by internal dynamics may be essential elements in
information processing \cite{Benyishai}.  Another deficiency of
all-to-all couplings is that real neurons are usually more sparsely
connected.  On the other hand, locally connected networks are free
from these shortcomings and can reproduce synchronization of different
neural assemblies sharing a preferred stimulus as far as the
assemblies are spatially close \cite{Konig,Wang95_IEEE}.  However,
slow information transmission via local connections can be an obstacle
to binding remote objects, and real networks are likely to have much
shorter pairwise distances than locally connected networks do.

Small-world networks \cite{Callaway_PRE,Watts,Wattsbook}
simultaneously realize dense local connections and short pairwise
distances. These small-world properties are also found in biological
neural networks \cite{Sporns,Stephan,Wattsbook}.  In this paper we
examine cooperative effects of local coupling, global coupling, and
coherent inputs on synchronization.

The ordinary small-world networks and the methodologies in this study
are explained in \SECS\ref{sec:smallworld} and \ref{sec:model},
respectively.  Then we point out in \SEC\ref{sec:dispersed} that the
ordinary small-world networks are not appropriate for studying
topological effects on dynamics of neural networks and introduce
modified small-world networks.  We next show in \SECS\ref{sec:feature}
and \ref{sec:gray} that small-world networks enable two remote groups
of neurons to synchronize when they receive coherent inputs.  It is
discussed in \SEC\ref{sec:discussion} that precise local synchrony and
rough global synchrony are bridged by the rewiring probability
specifying the randomness of networks.

\section{Small-world Networks}\label{sec:smallworld}

The characteristic path length $L$ and the clustering coefficient $C$
are commonly used to quantify the distances between vertices and the
clustering property, respectively \cite{Watts,Wattsbook}.  With a
graph composed of $n$ vertices, $L$ is defined to be the shortest path
length between two points averaged over all possible $n(n-1)/2$ pairs.
On the other hand, a vertex with $k_v$ neighbors could have at most
$k_v(k_v-1)/2$ edges between pairs of the neighbors. The number of
actual edges normalized by $k_v(k_v-1)/2$ and averaged over all the
vertices gives $C$. Many real networks including social, ecological,
and neural networks have small $L$ and large $C$, whereas regular
lattices have large $L$ and large $C$ and random sparse networks have
small $L$ and small $C$. Although all-to-all networks have small $L$
and large $C$, they have unrealistically many edges.

\begin{figure}
\centering
\includegraphics[width=0.49\textwidth]{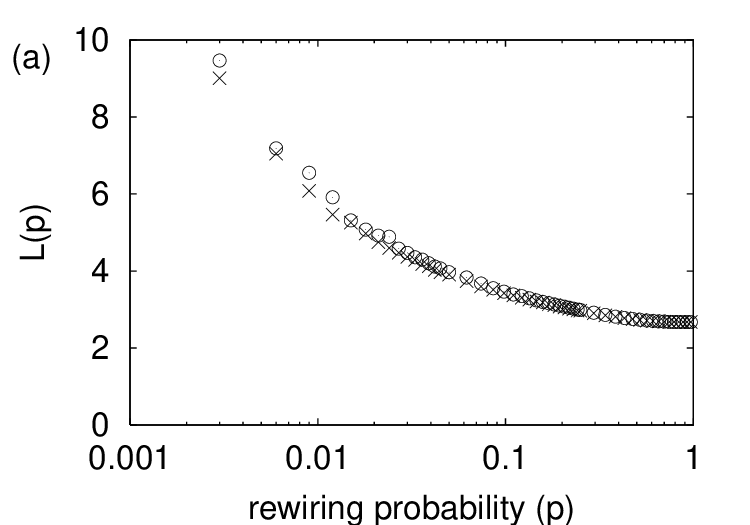}
\includegraphics[width=0.49\textwidth]{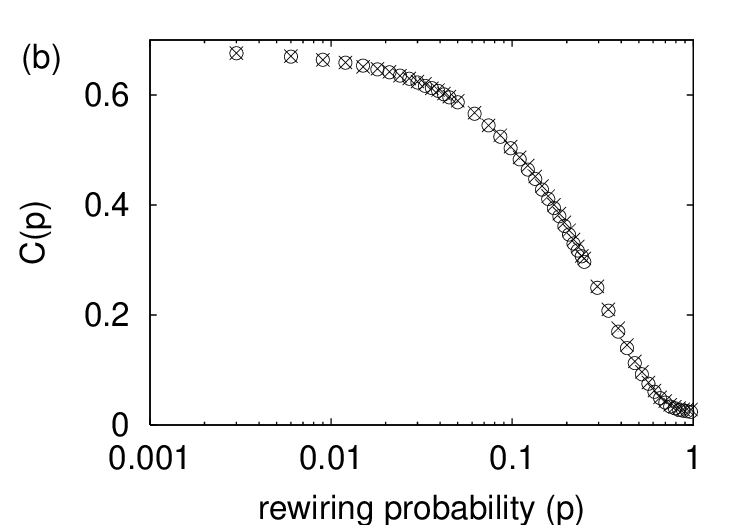}
\caption{(a) $L(p)$ and (b) $C(p)$ for $n=400$ and $k=6$. The graphs
are generated by the unbalanced random rewiring (crosses) and by the
balanced random rewiring (circles).}
\label{fig:graphstat}
\end{figure}

To generate small-world networks with small $L$ and large $C$, we
prepare a ring of $n$ vertices in which each vertex has $k/2$ nearest
neighbors on each side \cite{Watts,Wattsbook}.  Then we rewire a
fraction $p$ of edges randomly by removing $pkn/2$ edges and creating
$pkn/2$ new edges each of which leaves from the start point of a
removed edge and ends at a randomly chosen vertex, avoiding multiple
edges or recreation of the removed edges. As is shown in
\FIG\ref{fig:graphstat} for $n=400$ (crosses), rewired graphs with
$p\in[0.01,0.1]$ have the small-world properties with small $L(p)$ and
large $C(p)$.  The regular lattice ($p=0$) and the random graph
($p=1$) are generated by this procedure as two extremes
\cite{Watts,Wattsbook}.  Regarding neurobiology, the neural networks
of {\it C. elegans} \cite{Wattsbook} and the connections between the
cortex areas of macaque \cite{Sporns,Stephan} and cat \cite{Sporns}
are reported to be in the small-world regime.  Small-world networks
may serve to reduce wiring costs compared with the all-to-all
networks, to speed up information transmission compared with locally
connected networks, and to preserve spatial information on signals in
local assemblies in contrast to random networks.  Small-world effects
on synchronization have been explored with
pulse-coupled integrators \cite{Guardiola}, Hodgkin-Huxley neurons
\cite{Lagofernandez00,Lagofernandez01}, and FitzHugh-Nagumo neurons
\cite{Lagofernandez01}. However, the effect of heterogeneous $k_v$ for
nonzero $p$ and that of the connection structures have not been
distinguished.  Furthermore, the relation between network topology and
information processing with spatially heterogeneous inputs has not
been discussed except in \cite{Lagofernandez01} that deals with
localized inputs to trigger global synchrony and traveling waves.
Therefore, using modified small-world networks introduced in
\SEC\ref{sec:dispersed}, we focus on effects of coupling structure
on the dynamics with spatially structured
inputs. Following other theoretical works
\cite{Barahona,Guardiola,Lagofernandez00,Lagofernandez01,Sompolinsky91},
undirected graphs are used in this paper to facilitate the analysis
even though real neural systems seem to be directed graphs. The
synaptic connections are fixed once a network is specified.

\section{Model and Simulation Methods}\label{sec:model}

We use $n=400$ pulse-coupled leaky integrate-and-fire (LIF) neurons.
The $i$th LIF neuron ($1\le i\le n$) obeys the following dynamics:
\begin{equation}
\dot{x}_i(t) = -\gamma x_i(t) + I^{syn}_i(t) + I_i(t),
\end{equation}
where $x_i(t)$ is the membrane potential, and $\gamma$ is the leak
rate.  When $x_i(t)$ reaches 1, the neuron fires and $x_i(t)$ is reset
to the resting potential $x_i(t+0)=0$.  During the absolutely
refractory period $\tau^{\prime}=0.6$ ms after firing, the membrane
potential stays at the resting potential.  With the time delay $\tau$
after firing, each connected neuron receives an instantaneous spike
with the amplitude $\epsilon$.  These feedback interactions define the
synaptic input $I^{syn}_i(t)$.  The bias current $I_i(t)$ reflects
spatiotemporal features of external stimuli.  Although homogeneous
inputs $I_i(t)=I(t)$ for $1\le i\le n$ easily lead to stable full
synchrony \cite{Vanvreeswijk96} that is robust against some
heterogeneity \cite{Golomb00,Wang96}, sufficiently heterogeneous
$I_i(t)$, which we concentrate on, is more concerned with clustered
firing or local synchrony.

We define a synchrony measure $syn$ within a group of neurons ${\cal
S}\subset \{ 1, 2, \ldots, n\}$ \cite{Lagofernandez00} by
\begin{equation}
syn({\cal S}) = \sigma \left( \frac{1}{|{\cal S}|}
\sum_{i\in {\cal S}} x_i \right) \bigg/ 
\frac{1}{|{\cal S}|} \sum_{i\in {\cal S}} \sigma(x_i),
\end{equation}
where $|{\cal S}|$ is the cardinality of ${\cal S}$, and the
fluctuation $\sigma(x)$ of a temporal waveform $x$ up to time $T$ is
given by
\begin{equation}
\sigma(x)^2 = \frac{1}{T} \int^T_0 \left( x(t) - 
\frac{1}{T}\int^T_0 x(t^{\prime}) dt^{\prime} \right)^2 dt.
\end{equation}
Consequently, $syn$ is normalized between 0 and 1. More synchronous
firing gives larger values of $syn$ since the population average of
$x_i$ behaves more similarly to each $x_i$.  The values of $syn$ shown
in the following figures are the averages based on 25 runs.

We use the Euler algorithm with a time step equal to $0.05$ ms for
numerical simulations, except when examining robustness of the results
against different integration methods in \SEC\ref{sub:input_const}.

\section{Effect of Spread Vertex Degrees on Synchrony}\label{sec:dispersed}

Here the effect of heterogeneous vertex degrees is
examined.  In this section, we assume that the inputs are static:
$I_i(t)=I_i$, and that $I_i$ are distributed according to the uniform
distribution on $[I_0(1-\Delta_I), I_0(1+\Delta_I)]$ where $I_0 =
28.5$ and $\Delta_I = 0.08$.  Since $I_i>\gamma$, the neurons are
oscillatory.  We also set $k=12$, $\epsilon=0.05$, $\gamma = 25$
ms${}^{-1}$, and $\tau=0.5$ ms.

\begin{figure}
\centering
\includegraphics[width=0.49\textwidth]{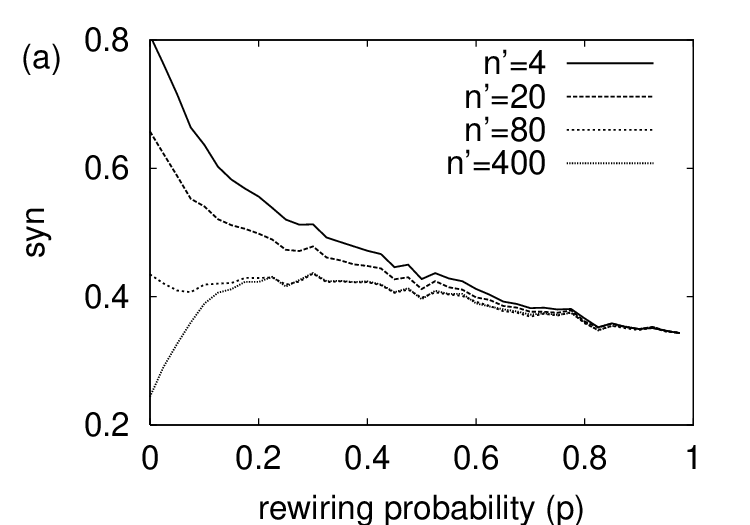}
\includegraphics[width=0.49\textwidth]{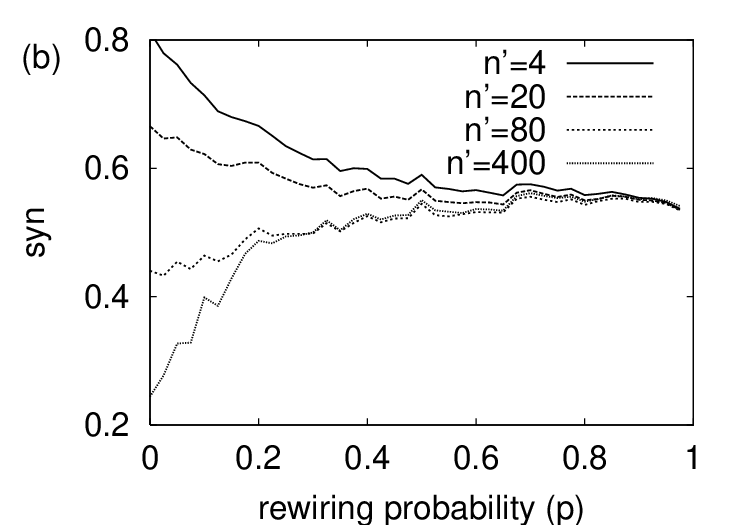}
\caption{The synchrony measure $syn$ with (a) the unbalanced random
rewiring, and (b) the balanced random rewiring. The size of local
groups of neurons within which $syn$ is evaluated is denoted by
$n^{\prime}$.}
\label{fig:un_bd}
\end{figure}

Figure~\ref{fig:un_bd}(a) shows how the conventional rewiring
(unbalanced rewiring) introduced in \SEC\ref{sec:smallworld}
influences the degree of synchrony.  To inspect synchrony on various
spatial scales, $syn({\cal S})$ averaged over all
possible ${\cal S}$ that consist of $n^{\prime}\le n$ adjacent neurons
on the ring substratum are displayed in \FIG\ref{fig:un_bd}(a).  The
best global synchrony ($n^{\prime}=400$, the bottom line) is obtained
with $p\cong 0.3$, although the peak is not so prominent.  This is
consistent with the results for coupled Hodgkin-Huxley neurons
\cite{Lagofernandez00}, whereas the optimal values of $p$ are much
larger here as for the results for coupled oscillators
\cite{Barahona}.  For smaller $p$, traveling waves or clustered states
appear.  In this situation, as the top line ($n^{\prime}=4$) in
\FIG\ref{fig:un_bd}(a) indicates, neurons locally synchronize owing to
the high intragroup connectivity characterized by large $C(p)$,
whereas large $L(p)$ prohibits global synchrony.  With intermediate
values of $p$, remote neurons communicate rapidly enough via shortcuts
to realize global synchrony, and abundant local connections guarantee
precise synchrony.  Too large values of $p$ result in asynchronous
firing.

However, \FIG\ref{fig:un_bd}(a) and other theoretical results
\cite{Barahona,Lagofernandez00,Lagofernandez01} do not necessarily
support biological relevance of small-world networks.  Related to
synaptic inputs, two desynchronizing factors are identified for
networks of homogeneous neurons \cite{Golomb00,Vanvreeswijk98},
namely, (i) spatially heterogeneous inputs, and (ii) heterogeneity in
the number of inputs to a neuron, or heterogeneous $k_v$.  The
positive values of $\Delta_I$ and the random rewiring underlie the
factors (i) and (ii), respectively.  When $p$ is large, the factor
(ii) overrides the tendency toward global synchrony caused by small
$L(p)$ \cite{Guardiola} and that toward local synchrony caused by
large $C(p)$, resulting in asynchrony.  As a result, the
enhanced synchronization with intermediate values of $p$ is not
entirely of the topological origin although small-world networks
approximate to real neural networks in terms of $L(p)$ and $C(p)$.
The vertex degree is homogeneous for $p=0$ because of the perfect
symmetry, whereas that of the random graph ($p=1$) has the Poisson
distribution with mean $k$ \cite{Erdos}.  Real neural networks would
not have homogeneous $k_v$ no matter whether they are close to
regular, small-world, or random networks. It is possible to construct
modified graph families whose degree distributions obey the Poisson
distribution independent of $p$ \cite{Callaway_PRE,Molloy}.  However,
here we treat more tractable cases with $k_v=k$ for all the vertices
regardless of $p$ because synchrony is more easily attained with more
homogeneous networks, especially when networks are small as used in
numerical simulations.

Figure~\ref{fig:graphstat} compares $L(p)$ and $C(p)$ of the graphs
based on the original rewiring procedure (unbalanced random rewiring)
and the modified rewiring procedure (balanced random rewiring).  To
generate balanced networks, we first remove $pkn/2$ edges whose
endpoints are denoted by $v_{i,1}$ and $v_{i,2}$ ($1\le i\le pkn/2$)
as is done in generating unbalanced networks. Let us denote $\rho$ a
randomly chosen permutation on $\{1, 2, \ldots, pkn/2\}$. Then new
edges are created by connecting $v_{i,1}$ and $v_{\rho(i),2}$ ($1\le
i\le pkn/2$). Another permutation is tried if there appear multiple
edges, a removed edge, or self-connection.  Insusceptibility of $L(p)$
and $C(p)$ to the graph modification evident in
\FIG~\ref{fig:graphstat} permits us to explore topological effects on
dynamics using the balanced random rewiring.

Degrees of synchrony with the unbalanced and balanced rewiring are
compared in \FIGS\ref{fig:un_bd}(a) and (b). Figure~\ref{fig:un_bd}(b)
shows that the optimal $p$ for synchrony vanishes with the balanced
rewiring.  Synchrony is induced in original small-world networks when
$L(p)$ and the heterogeneity in $k_v$ are simultaneously small. In the
balanced networks, however, effects of heterogeneous $k_v$ are
abolished.  Now, global synchrony just improves as $p$ increases
($n^{\prime}=400$, the bottom line) since the contribution of small
$L(p)$ in favor of global synchronization surpasses that of small
$C(p)$ in favor of desynchronization.  By the same token, diffusively
coupled neurons do not yield the optimal $p$ in the small-world regime
even in the unbalanced networks \cite{Wattsbook}. This is because the
influence of heterogeneous $k_v$ in diffusively coupled networks is
less prominent than in pulse-coupled cases.

Locally, rapid communication between pulse-coupled neurons is possible
irrespective of $L(p)$ and $p$.  Local synchrony degrades as $p$
increases (see $n^{\prime}=4$, the top line) since $C(p)$ rather than
$L(p)$ determines the possibility of local synchronization.

\section{Synchrony of Remote Neurons with Coherent Inputs}\label{sec:feature}

Here we shed light on how the network structure affects local and
global synchrony of remote neuronal assemblies. 
The combined effects of network topology
and coherence of inputs to different neurons
\cite{Benyishai,Gray,Sompolinsky90,Sompolinsky91} are also detailed.

\subsection{Constant Bias Inputs}\label{sub:input_const}

\begin{figure}
\centering
\includegraphics[width=0.49\textwidth]{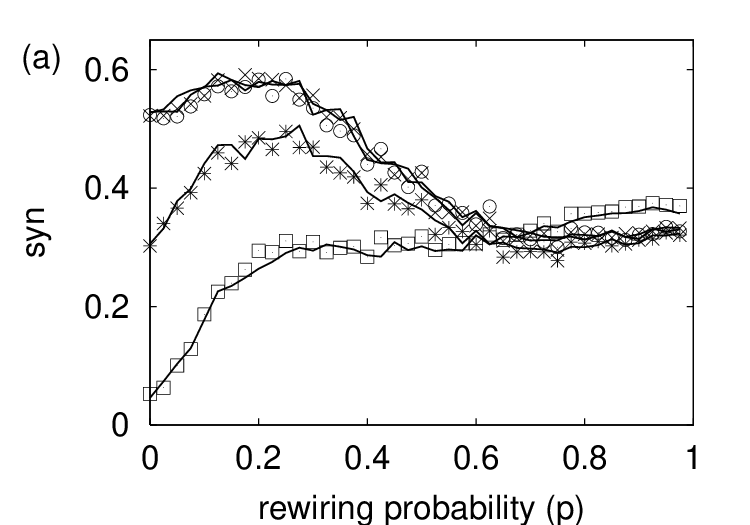}
\includegraphics[width=0.49\textwidth]{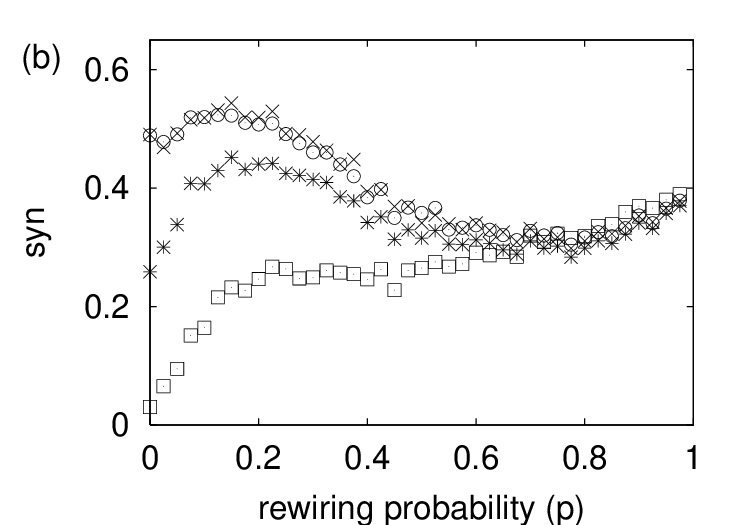}
\includegraphics[width=0.49\textwidth]{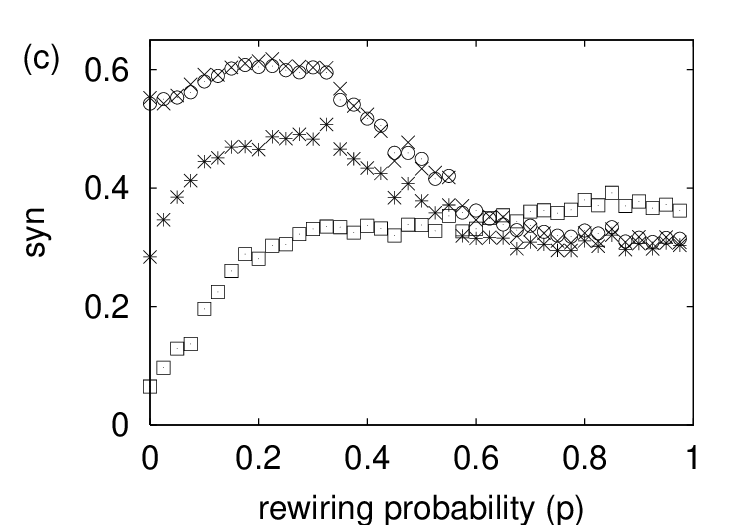}
\includegraphics[width=0.49\textwidth]{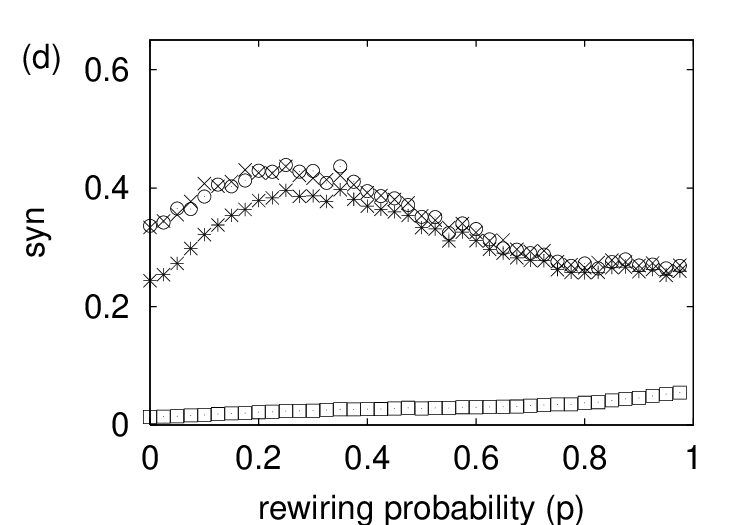}
\caption{The synchrony measure $syn({\cal S}_1)$ (crosses), $syn({\cal
S}_2)$ (circles), $syn({\cal S}_1\cup {\cal S}_2)$ (asterisks), and
$syn({\cal S}_1\cup {\cal S}_2\cup \overline{\cal S})$ (squares) when
two separated coherent stimuli are presented. (a), (b), and (c) are
the results for bias inputs with $(k,\epsilon)=(12,0.05)$,
$(8,0.073)$, and $(18,0.035)$, respectively.  (d) shows the results
for correlated noise inputs with $(k,\epsilon)=(12,0.05)$.  The
results are obtained by the normal Euler algorithm except those
indicated by the solid lines in (a), which are obtained by the
improved Euler algorithm.}
\label{fig:twobars}
\end{figure}

We first deal with temporally constant and spatially structured
inputs.  In relation to feature binding as we discuss in
\SEC\ref{sub:feature}, let us assume that two separated neuronal
assemblies receive a common bias added to the background level.  The
assemblies ${\cal S}_1$ and ${\cal S}_2$ are located on the opposite
sides of the ring: $I_i = 1.1 I_0$ for $i\in {\cal S}_1\cup
{\cal S}_2$, ${\cal S}_1 = \{ i ; 0 < i\le 0.1n \}$, and ${\cal S}_2 =
\{ i ; 0.5n < i \le 0.6n\}$.  The background activity $I_i$ given to
the neurons $\overline{\cal S} = \{1,2, \ldots, n\} \backslash ({\cal
S}_1\cup {\cal S}_2)$ is taken randomly from the uniform distribution
on $[I_0(1-\Delta_I),I_0(1+\Delta_I)]$.  We put $\epsilon=0.05$,
$\gamma=40$ ms${}^{-1}$, $I_0=38.5$ ms${}^{-1}$, $\Delta_I = 0.1$,
$\tau=1.2$ ms, and uncorrelated Gaussian white noise with the standard
deviation $1.0\times 10^{-3}$ per step is applied to all the neurons.
Figure~\ref{fig:twobars}(a) shows degrees of synchrony of interest,
namely, $syn({\cal S}_1)$, $syn({\cal S}_2)$, $syn({\cal S}_1\cup
{\cal S}_2)$, and $syn({\cal S}_1\cup {\cal S}_2\cup \overline{\cal
S})$.  Local precise synchrony within ${\cal S}_1$ or ${\cal S}_2$
requires large $C(p)$ associated with small $p$, with synchrony
deteriorating as $p$ increases.  On the other hand, the intergroup
synchrony is enhanced when $p\in [0.2,0.5]$, with $syn({\cal S}_1\cup
{\cal S}_2)$ significantly larger than $syn({\cal S}_1\cup {\cal
S}_2\cup \overline{\cal S})$. Intergroup synchrony is difficult when
$p$ is too small because large $L(p)$ prohibits ${\cal S}_1$ and
${\cal S}_2$ from communicating fast enough.
Intermediate values of $p$ let remote groups of neurons
synchronize with the help of shortcuts while local connections
necessary to maintain precise intragroup synchrony are still
abundant. Although rough global synchrony emerges with larger $p$,
synchrony of the neurons receiving the common inputs is deteriorated
because the loss of local connectivity forces these neurons to
interact with a random selection of neurons receiving heterogeneous
inputs \cite{Wang96}.

For $(k,\epsilon) = (8,0.073)$ and $(k,\epsilon) = (18,0.035)$ with
the other parameter values fixed, the values of $syn$ are shown in
\FIGS\ref{fig:twobars}(b) and (c), respectively. The firing rates are
kept unchanged by modulating $\epsilon$.  The values of $p$ for
improved intergroup synchrony are robust with respect to $k$ and are
larger than those for the small-world regime \cite{Barahona}.  The
clustering property measured by $C(p)$ is based on the number of
connected triangles in the network.  Therefore, even though $C(p)$ is
small with $p\cong 0.3$ as demonstrated in \FIG\ref{fig:graphstat}(b),
the local edges within ${\cal S}_1$ or within ${\cal S}_2$ are still
abundant enough for intragroup synchrony.  Given intragroup synchrony,
what promotes intergroup synchrony is the global connections between
${\cal S}_1$ and ${\cal S}_2$. However, a small value of $L(p)$
reminiscent of the small-world property does not necessarily indicate
sufficient global connections. When $p$ is small, the number
of global connections increases linearly with $p$, and $L(p)$
decreases logarithmically with $p$ \cite{Watts}. Actually, $p$ must be
larger than for small-world networks to ensure effective interactions
between remote assemblies.  In spite of the deviation of the optimal
range for $p$ from the small-world regime, improved intergroup
synchrony is realized only when the local and global interactions
cooperate.

As a remark, numerical results for LIF neurons are generally sensitive
to the simulation time step unless firing time is calculated by the
appropriate interpolation that the standard Euler algorithm has
ignored \cite{Hansel98}. The lines in \FIG\ref{fig:twobars}(a)
indicate values of $syn$ when the improved Euler algorithm
\cite{Hansel98} is implemented. The results do not critically depend
on the integration methods, which implies the robustness of the
results. This is also true for other numerical results in this paper.

\subsection{Correlated Noise Inputs}\label{sub:input_ou}

Constant inputs are often too simple to mimic real inputs
with a large amount of fluctuation.  Additionally, neurons receiving
coherent temporally modulated inputs are more likely to synchronize
than when receiving coherent constant inputs \cite{Mainen}.  Here we
use the Ornstein-Uhlenbeck processes to emulate such inputs.  The
Ornstein-Uhlenbeck process \cite{Mainen,Masuda_DUAL} is defined by
\begin{equation}
\frac{dy_i(t)}{dt} =  - \frac{y_i(t)}{\tau_{ou}} + \xi_i(t),
\label{eq:input_ou1}
\end{equation}
where we set $\tau_{ou} = 3.0$ ms. Gaussian white noise $\xi_i(t)$ is
applied at every time step to emulate the Brownian motion. The spatial
inputs simulating two coherent but remote input assemblies are defined
by
\begin{equation}
I_i(t) = I_0 + 0.09 \; y_i(t), \quad (1\le i\le n),
\label{eq:input_ou2}
\end{equation}
where the correlation between $\xi_i(t)$ and $\xi_j(t)$ is 0.8 if
$i,j\in {\cal S}_1 \cup {\cal S}_2$ and 0 otherwise.  Figure
\ref{fig:twobars}(d) shows the numerical results for the
Ornstein-Uhlenbeck inputs. The parameters take the same values as
those for \FIG\ref{fig:twobars}(a) except that we set $I_0=0.044$ to
preserve the firing rates. Enhanced synchrony with intermediate values
of $p$ is also observed in this case.  A remarkable difference from
\FIG\ref{fig:twobars}(a) is that the neurons are globally asynchronous
even when $p$ is large.  Only the neurons receiving correlated inputs
are capable of maintaining synchrony, and the difference in $syn({\cal
S}_1 \cup {\cal S}_2)$ and $syn({\cal S}_1 \cup {\cal S}_2\cup
\overline{\cal S})$ is ascribed to the correlated inputs
\cite{Mainen}. The additional amount of $syn$ when $p\cong 0.3$ owes
to global coupling.  As a remark, $syn({\cal S}_1)$ and $syn({\cal
S}_2)$ significantly decrease for small $p$ compared with the case of
constant inputs. This might be because of the relatively short range
of the local connections.

\section{Degree of Synchrony Modulated by Input Structure}\label{sec:gray}

We have seen that distant neurons fire synchronously only when they
are stimulated by the coherent constant or coherent noisy inputs.  In
experiments with visual stimuli, even better synchrony takes place
when two remote but coherent inputs are interpreted to be united as
one global stimulus \cite{Gray}. Related to these experimental facts,
we examine how inputs with
different spatial structure affect spatial firing patterns.

\begin{figure}
\centering
\includegraphics[width=0.49\textwidth]{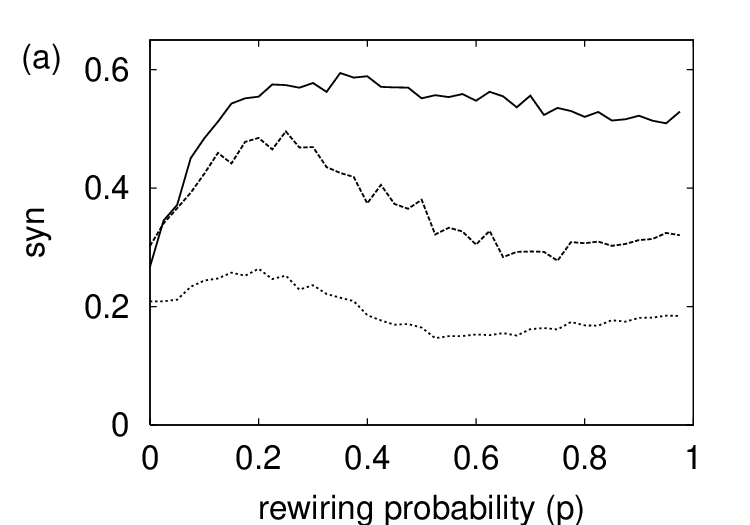}
\includegraphics[width=0.49\textwidth]{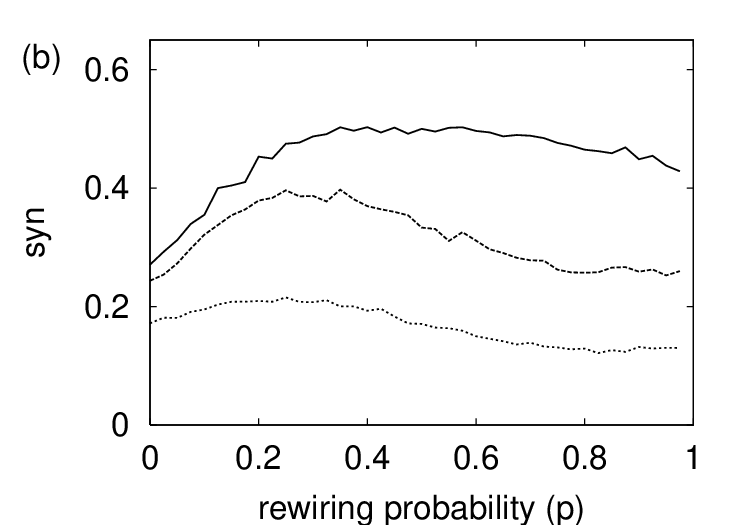}
\caption{The synchrony measure $syn({\cal S}_1\cup {\cal S}_2)$ when
(I) a globally coherent input (solid lines), (II) two separated
coherent inputs (dashed lines), and (III) two separated incoherent
inputs (dotted lines) are presented.  The inputs are applied as (a)
biases or (b) correlated noise. We set $k=12$ and $\epsilon=0.05$.}
\label{fig:gray}
\end{figure}

First, three types of bias inputs, namely, (I) a globally coherent
input $I_i = 1.1 I_0$ for $0< i\le 0.6n$, (II) two separated coherent
inputs $I_i=1.1 I_0$ for $i\in {\cal S}_1\cup {\cal S}_2$, and (III)
two locally coherent but globally incoherent inputs $I_i=1.1 I_0$ for
$i\in {\cal S}_1$ and $I_i = 1.15 I_0$ for $i\le {\cal S}_2$, are
applied. Input (II) is equivalent to one used in
\SEC\ref{sub:input_const}.  The bias $I_i$ to the neurons in the
background and the other parameter values are the same as before.  As
shown in \FIG~\ref{fig:gray}(a), $syn({\cal S}_1\cup {\cal S}_2)$
increases with input (I) and decreases little with an increasing $p$
since the diminishing local communications are compensated by the
global connections created by random rewiring, many of which still
fall within the range of the global input.  On the other hand, the
neurons are much less synchronized with input (III) since the global
coupling cannot compensate the desynchronizing effects due to the
heterogeneous inputs to ${\cal S}_1$ and ${\cal S}_2$ \cite{Wang96}.

The corresponding results with the correlated noise inputs introduced
in \SEC\ref{sub:input_ou} are shown in \FIG\ref{fig:gray}(b).  We
assume that the correlation between $I_i$ and $I_j$ is 0.8 only when
inputs to the $i$th and $j$th neurons are supposed to be coherent.
The difference between $syn$ for input (II) and that for (III), which
depends little on $p$, is attributed to the correlated inputs. The
additional amount of $syn$, which is nearly equal to the $syn$ value
for input (III), is induced by the global coupling. This result
supports that synchrony is reinforced with intermediate $p$ purely for
the topological reason.  Input (I) realizes larger $syn$ as expected.

\section{Discussion}\label{sec:discussion}

\subsection{Two Modes of Dynamics}\label{sub:spatial_dual}

Figure~\ref{fig:un_bd}(b) shows a tradeoff between the temporal
precision and the spatial extent of synchrony. For small $p$, local
precise synchrony appears (see the top line), whereas neurons are
globally asynchronous (see the bottom line). In this regime, many
locally synchronous clusters coexist and can encode inputs with higher
spatial resolution. At the extreme, each neuron fires independently of
all the others to represent the corresponding $I_i(t)$. By contrast,
for large $p$, $syn$ is almost independent of the group size
$n^{\prime}$ since physical locations of neurons do not make sense;
spatiotemporal inputs are
encoded crudely in the sense of spatial resolution but robustly.  As
$p$ increases, rough global synchrony replaces local precise
synchrony.

These results contrast with the dual coding in feedforward networks
that receive a temporal signal
\cite{Masuda_IEI,Masuda_DUAL,Vanrossum02}. With strong coupling, a
small amount of noise, and homogeneity, neurons synchronize,
possibly resorting to rough but power-saving synchronous coding with
spatiotemporal cell assemblies \cite[pp. 141--173]{Domany}. This mode
is a generalization of the synfire chains \cite{Abeles91},
\cite[pp. 121--140]{Domany}.  With weak coupling, a large amount of
noise, and heterogeneity, neurons desynchronize, possibly encoding
temporal signals more precisely by the population rate coding.  On the
other hand, recurrent networks with spatially structured inputs are
investigated in this paper.  Robust coding with global synchrony and
spatially accurate coding with rough precise 
local synchrony are in the tradeoff
relation. The brain may modulate $p$ to switch between the two modes
through synaptic plasticity
\cite{Domany,Masuda_DUAL,Vanvreeswijk96_SCI,Vanvreeswijk98}.

\subsection{Feature Binding}\label{sub:feature}

In experiments, when two visual objects moving in the same direction
are presented, neurons in the visual cortex receiving the same
preferred stimuli fire synchronously even if the neurons are spatially
separated \cite{Gray}. Since the topological structure of receptive
fields generally agrees with the anatomical one for example in the
primary visual cortex, current biases higher than the background level
are often given to two separated groups of neurons to model two
objects.  With local connections only, nearby objects can be united as
a synchronous cluster by intergroup interaction, but presentation of
remote objects does not allow global intergroup
synchrony \cite{Konig,Wang95_IEEE}. More elaborated networks with
local excitations and global uniform inhibitions realize cluster
coding in which different objects are encoded by separated synchronous
groups that fire alternatively \cite{Terman,Vondermalsburg86}.

We have shown that the cooperation of the local and global
interactions, which is realized by intermediate values of $p$,
facilitates synchrony among remote groups of neurons.  The
combined effects of couplings and coherent inputs have also been
emphasized.  The results may be relevant to binding of remote coherent
stimuli such as visual object segmentation \cite{Terman,Wang96}, the
synfire chain \cite{Abeles91}, and dynamical cell assembly
\cite{Fujii}, although we have ignored the diversity of the stimulus
preferences of individual neurons such as orientation selectivity and
simply represented a coherent object as a common signal with a spatial
structure.  Also in terms of network topology, our results are
consistent with the findings that nearby neurons with overlapping
receptive fields are densely and strongly connected irrespective of
preferred stimuli, while weak long-range connections exist only
between the neurons with the common preferred stimuli
\cite{Sompolinsky90}.

Another point to be noted is that the occurrence of synchrony is
unrelated to the locations of external inputs as long as they are
coherent. Other model studies of feature binding
require closeness of objects \cite{Konig,Wang95_IEEE}, homogeneous
global inhibitors \cite{Terman,Vondermalsburg86}, or the
stimulus-dependent synaptic strength that changes rapidly
\cite{Sompolinsky90,Sompolinsky91,Vondermalsburg86}.  Our neural
networks with the cooperative local and global connections do not rely
on these requirements that may be biologically too strong.

\subsection{On Plausibility of Using Small-world Networks}

With our results, we do not insist that small-world networks naturally
emerges to facilitate synchrony or information processing.  As
\FIG\ref{fig:gray} shows, separation of three kinds of inputs (I),
(II), and (III) in \SEC\ref{sec:gray} is also achieved with larger
$p$. The purpose in the analysis in \SECS\ref{sec:feature} and
\ref{sec:gray} was to examine the topological effect on synchrony and
manifest the cooperation of couplings and inputs.  Randomly rewired
networks including small-world networks are used just for the same
goal.  For example, synfire chains \cite{Abeles91},
\cite[pp. 121--140]{Domany} in which precise local synchrony as well
as global information transmission is essential may be related to
rewired networks with intermediate values of $p$
\cite{Lagofernandez00}.  However, random rewiring always discards
topological information.  Real neural networks may have particular
structures and inherent correlations as a result of learning even if
$L(p)$ and $C(p)$ resemble those of randomly shuffled small-world networks
\cite{Callaway_PRE}.  To consider learning, dynamics of synaptic
connectivities, and also inhibitory couplings is an important future
problem.

\section*{Acknowledgements}
We thank M. Watanabe, Y. Tsubo, and C. van Leeuwen for helpful
discussions.  This work is partially supported by the Japan Society
for the Promotion of Science and by the Advanced and Innovational
Research program in Life Sciences and a Grant-in-Aid No.15016023 on
priority areas (C) Advanced Brain Science Project from the Ministry of
Education, Culture, Sports, Science, and Technology, the Japanese
Government.


\begin{thebibliography}{000}

\bibitem{Abeles91}
Abeles M (1991)
Corticonics. Cambridge University Press, Cambridge.

\bibitem{Barahona}
Barahona M, Pecora LM (2002) Synchronization in small-world systems.
Phys Rev Lett 89(5): 054101

\bibitem{Benyishai}
Ben-Yishai R, Lev Bar-Or R, Sompolinsky H (1995) Theory of
orientation tuning in visual cortex. 
Proc Natl Acad Sci USA 92: 3844--3848

\bibitem{Callaway_PRE}
Callaway DS, Hopcroft JE, Kleinberg JM, Newman MEJ, Strogatz SH (2001)
Are randomly grown graphs really random? Phys Rev E 64: 041902.

\bibitem{Domany}
Domany E, van Hemmen JK, Schulten K (Eds.) (1994) Models of neural
networks II. Springer-Verlag, New York.

\bibitem{Erdos}
Erd\"{o}s P, R\'{e}nyi A (1959)
On random graphs.
Publicationes Mathematicae 6: 290--297.

\bibitem{Fujii}
Fujii H, Ito H, Aihara K, Ichinose N, Tsukada M (1996)
Dynamical cell assembly hypothesis -- theoretical
possibility of spatio-temporal coding in the cortex.
Neural Netw 9(8): 1303--1350

\bibitem{Golomb00}
Golomb D, Hansel D (2000)
The number of synaptic inputs and the synchrony of large, sparse
neuronal networks.
Neural Comput 12: 1095--1139.

\bibitem{Gray}
Gray CM, K\"{o}nig P, Engel AK, Singer W (1989)
Oscillatory responses in cat visual cortex exhibit
inter-columnar synchronization which reflects global stimulus
properties.
Nature 338: 334--337

\bibitem{Guardiola}
Guardiola X, D\'{\i}az-Guilera A, Llas M, P\'{e}rez CJ (2000)
Synchronization, diversity, and topology of integrate and fire
oscillators.
Phys Rev E 62(4): 5565--5570

\bibitem{Hansel98}
Hansel D, Mato G, Meunier C, Neltner L (1998)
On numerical simulations of integrate-and-fire neural networks.
Neural Comput 10: 467--483

\bibitem{Konig}
K\"{o}nig P, Schillen TB (1991)
Stimulus-dependent assembly formation of oscillatory responses:
I. synchronization.
Neural Comput 3: 155--166

\bibitem{Lagofernandez00}
Lago-Fern\'{a}ndez LF, Huerta R, Corbacho F,
Sig\"{u}enza JA (2000)
Fast response and temporal coherent oscillations
in small-world networks. 
Phys Rev Lett 84(12):
2758--2761

\bibitem{Lagofernandez01}
Lago-Fern\'{a}ndez LF, Corbacho FJ, Huerta R (2001)
Connection topology dependence of synchronization of neural
assemblies on class 1 and 2 excitability.
Neural Netw 14: 687--696

\bibitem{Mainen}
Mainen ZF, Sejnowski TJ (1995)
Reliability of spike timing in neocortical neurons
Science 268: 1503--1506

\bibitem{Masuda_IEI}
Masuda N, Aihara K (2002) Bridging rate coding and
temporal spike coding by effect of noise. Phys Rev Lett
88(24): 248101

\bibitem{Masuda_DUAL}
Masuda N, Aihara K (2003)
Duality of rate coding and temporal coding in multilayered
feedforward networks. 
Neural Comput 15: 103--125 

\bibitem{Molloy}
Molloy M, Reed B (1995) A critical point for random graphs with a
given degree sequence.
Random Struct Algorithms 6: 161--180

\bibitem{Sompolinsky90}
Sompolinsky H, Golomb D, Kleinfeld D (1990)
Global processing of visual stimuli in a neural network of coupled
oscillators. 
Proc Natl Acad Sci USA 87: 7200--7204

\bibitem{Sompolinsky91}
Sompolinsky H, Golomb D, Kleinfeld D (1991)
Cooperative dynamics in visual processing.
Phys Rev A 43(12): 6990--7011

\bibitem{Sporns}
Sporns O, Tononi G, Edelman GM (2000)
Theoretical Neuroanatomy:
relating anatomical and functional connectivity in graphs and cortical
connection matrices. 
Cerebral Cortex 10: 127--141.

\bibitem{Steinmetz}
Steinmetz PN, Roy A, Fitzgerald PJ, Hsiao SS,
Johnson KO,  Niebur E (2000)
Attention modulates
synchronized neuronal firing in primate somatosensory cortex.
Nature 404: 187--190

\bibitem{Stephan}
Stephan KE., Hilgetag C-C, Burns GAPC,
O'Neill MA, Young MP, K\"{o}tter R (2000)
Computational analysis of functional
connectivity between areas of primate cerebral cortex.
Phil Trans R Soc Lond B 355: 111--126

\bibitem{Terman}
Terman D, Wang D (1995)
Global competition and local cooperation in a network of neural
oscillators.
Physica 81D: 148--176

\bibitem{Vanrossum02}
van Rossum MCW, Turrigiano GG, Nelson SB (2002) Fast
propagation of firing rates through layered networks of noisy
neurons. J Neurosci 22(5): 1956--1966

\bibitem{Vanvreeswijk96}
van Vreeswijk C (1996)
Partial synchronization in populations of
pulse-coupled oscillators.
Phys Rev E 54(5): 5522--5537

\bibitem{Vanvreeswijk96_SCI}
van Vreeswijk C, Sompolinsky H (1996)
Chaos in neuronal networks with
balanced excitatory and inhibitory activity.
Science 274: 1724--1726

\bibitem{Vanvreeswijk98}
van Vreeswijk C, Sompolinsky H (1998)
Chaotic balanced state in a model of cortical circuits.
Neural Comput 10: 1321--1371

\bibitem{Vondermalsburg86}
von der Malsburg Ch, Schneider W (1986)
A neural cocktail-party processor.
Biol Cybern 54: 29--40

\bibitem{Wang95_IEEE}
Wang D (1995)
Emergent synchrony in locally coupled neural oscillations. 
IEEE Trans on Neural Netw 6(4): 941--948

\bibitem{Wang96}
Wang XJ, Buzs\'{a}ki G (1996)
Gamma oscillation by synaptic inhibition in a hippocampal
interneuronal network model.
Journal of Neurosci 26(20): 6402--6413

\bibitem{Watts}
Watts DJ, Strogatz SH (1998)
Collective dynamics of `small-world' networks. 
Nature 393: 440--442

\bibitem{Wattsbook}
Watts DJ (1999) Small worlds. Princeton University Press,
Princeton

\end{thebibliography}
\end{document}